\documentstyle[12pt,a4wide]{article}
\begin{document}
\input epsf
\newcommand{\be}{\begin{equation}}
\newcommand{\ee}{\end{equation}}
\newcommand{\bea}{\begin{eqnarray}}
\newcommand{\eea}{\end{eqnarray}}
\newcommand{\ti}{\tau^{-1}}
\newcommand{\pr}{\partial}
\newcommand{\ie}{{\it ie }}
\font\mybb=msbm10 at 11pt
\font\mybbb=msbm10 at 17pt
\def\bb#1{\hbox{\mybb#1}}
\def\bbb#1{\hbox{\mybbb#1}}
\def\Z {\bb{Z}}
\def\R {\bb{R}}
\def\E {\bb{E}}
\def\T {\bb{T}}
\def\C {\bb{C}}
\def\bP {\bb{P}}
\def\CP{\C\bP}
\renewcommand{\theequation}{\arabic{section}.\arabic{equation}}
\newcommand{\news}{\setcounter{equation}{0}}
\newcommand{\bx}{{\bf x}}
\newcommand{\half}{\frac{1}{2}}
\newcommand{\G}{{\cal G}}
\renewcommand{\tilde}{\widetilde}
\title{\vskip -70pt
\begin{flushright}
\end{flushright}\vskip 50pt
{\bf \large \bf INSTANTONS AND THE BUCKYBALL}\\[30pt]
\author{Paul Sutcliffe
\\[10pt]
\\{\normalsize   {\sl Institute of Mathematics,}}
\\{\normalsize {\sl University of Kent,}}\\
{\normalsize {\sl Canterbury, CT2 7NZ, U.K.}}\\
{\normalsize{\sl Email : P.M.Sutcliffe@kent.ac.uk}}\\}}
\date{September 2003}
\maketitle

\begin{abstract}
\noindent 
The study of Skyrmions predicts that there is an icosahedrally
symmetric charge seventeen $SU(2)$ Yang-Mills instanton in which the 
 topological charge density, for fixed Euclidean time, is localized
on the edges of the truncated icosahedron of the buckyball.
In this paper the existence of such an instanton is proved by explicit
construction of the associated ADHM data. A topological charge density 
isosurface is displayed which verifies the buckyball structure 
of the instanton.  

\end{abstract}
\newpage

\section{Introduction}\news\label{sec-intro}
\hskip 0.5cm
Skyrmions, which are three-dimensional topological solitons, have an
approximate description in terms of four-dimensional $SU(2)$ 
Yang-Mills instantons \cite{AM}. In this approach a charge $N$ Skyrme field 
is approximated by the holonomy, along lines parallel to the Euclidean time
axis, of a charge $N$ instanton in $\R^4.$
A rotational symmetry of a Skyrmion in $\R^3$ 
corresponds to an equivalent rotational symmetry of the instanton, acting
 as a rotation of $\R^3\subset \R^4$ leaving fixed the Euclidean time.

It is expected that all minimal energy Skyrmions (and  other non-minimal
Skyrme fields) can be adequately described by the instanton approximation.
Thus, if the minimal energy charge $N$ Skyrmion is
symmetric under the action of a finite rotation group $\G\subset SO(3),$
the instanton approximation predicts the existence of a (family of)
charge $N$ $\G$-symmetric instantons. The minimal energy Skyrmions of
charge one, two, three, four and seven are particularly symmetric,
having spherical, axial, tetrahedral, octahedral and icosahedral symmetry
respectively \cite{BTC,BS2}, and suitable symmetric instantons have
been found \cite{AM,LM,SiSu} to correspond to each of these.

For larger values of the charge the minimal energy Skyrmion
generically has a fullerene-like structure \cite{BS}, in which
the topological charge density is localized around the edges
of a trivalent fullerene polyhedron. It is therefore expected that
there are families of fullerene-like instantons, in which the 
instanton topological charge density, for fixed Euclidean time, is localized
on the edges of the fullerene polyhedron. A particularly symmetric example
occurs at charge seventeen, where the fullerene is the icosahedrally 
symmetric buckyball of the truncated icosahedron. Given that this
corresponds to the minimal energy charge seventeen Skyrmion then the
prediction is that there is an icosahedrally
symmetric charge seventeen Yang-Mills instanton in which the 
 topological charge density, for fixed Euclidean time, is localized
on the edges of the buckyball. In this paper we prove the existence of
such an instanton by explicit construction of its ADHM data.

The ADHM construction, which we briefly review in the
following section, converts the instanton equations into
nonlinear algebraic constraints. However, only for instantons with
charge three or less can the general solution of these constraints be
obtained in closed form. The construction of high charge symmetric
instantons, motivated by the existence of associated Skyrmions, may
therefore be viewed as a way to simplify the ADHM constraints so that
particular exact solutions may be found even though the general
solution is not tractable. For most symmetric instantons obtained
this way (including the one presented in this paper) elementary
symmetry considerations show that the instanton is not of the
Jackiw-Nohl-Rebbi type \cite{JNR}, so it is a genuinely
new solution of the ADHM constraints. 

\newpage
\section{Symmetric ADHM Data}\news\label{sec-ADHM}\hskip 0.5cm
The ADHM construction \cite{ADHM,CWS,CFGT} 
generates the gauge potential of
the general charge $N$ instanton from matrices satisfying certain
algebraic, but nonlinear, constraints. 

The ADHM data 
for an $SU(2)$ $N$-instanton consists of a matrix
\be
\widehat M=\pmatrix{L \cr M}
\label{adhmdecomp}
\ee
where $L$ is a row of $N$ quaternions and $M$ is a symmetric $N\times N$
matrix of quaternions. 

To be valid ADHM data the matrix $\widehat M$ must satisfy the nonlinear
reality constraint
\be
\widehat M^\dagger\widehat M=R_0\,,
\label{adhmcon}
\ee
where ${}^\dagger$ denotes the quaternionic conjugate transpose and $R_0$
is any
real non-singular $N\times N$ matrix. 

The first step in constructing the instanton from the ADHM data is to form
the matrix
\be
\Delta(x)=\pmatrix{L \cr M-x1_N }\,,
\label{adhmop}
\ee
where $1_N$ denotes the $N\times N$ identity matrix and $x$ is the
quaternion corresponding to a point in $\R^4$ via $x=x_4+ix_1+jx_2+kx_3$.
The second step is then to find the $(N+1)$-component column vector
$\Psi(x)$ of unit length, $\Psi(x)^\dagger \Psi (x)=1$, which solves
the equation
\be
\Psi(x)^\dagger\Delta(x)=0\,.
\label{adhmlin}
\ee
The final step is to compute the gauge potential $A_\mu(x)$ from $\Psi(x)$
using the formula
\be
A_\mu(x)=\Psi (x)^\dagger\partial_\mu \Psi(x)\,.
\label{adhmpot}
\ee
This defines a pure quaternion which can be regarded as an element of
$su(2)$ using the standard representation of the quaternions in terms
of the Pauli matrices.

In order for all these steps to be valid, the ADHM data must satisfy an
invertibility condition, which is that the columns of $\Delta(x)$
span an $N$-dimensional quaternionic space for all $x$. In other words,
\be
\Delta(x)^\dagger\Delta(x)=R(x)
\label{rmatrix}
\ee
where $R(x)$ is a real $N\times N$ invertible matrix for every $x$.

It will be useful later to recall that the topological charge density
\be
{\cal N}=-\frac{1}{16\pi^2}\epsilon_{\mu\nu\alpha\beta}\mbox{Tr}
(F_{\mu\nu}F_{\alpha\beta})
\ee
(whose integral over $\R^4$ gives the instanton number $N$)
 can be written entirely in terms
of the determinant of the matrix $R(x)$ as \cite{CFGT,Os}
\be
{\cal N}=-\frac{1}{16\pi^2} \nabla^2\nabla^2 \log{\rm det} R(x)
\label{neat}\ee
where $\nabla^2$ denotes the four-dimensional Laplacian.

There is a freedom in choosing $\Psi(x)$ given by
$\Psi(x)\mapsto \Psi (x)q(x)$,
where $q(x)$ is a unit quaternion.
The unit quaternions can be identified with $SU(2)$
and from equation (\ref{adhmpot}) we see that this freedom corresponds
to a gauge transformation.

There is a further redundancy in the ADHM data
corresponding to the transformation
\be
\Delta(x)\mapsto \pmatrix{q & 0 \cr 0 & {\cal O}}\Delta(x){\cal O}^{-1}\,,
\label{adhmred}
\ee
where ${\cal O}$ is a constant
real orthogonal $N\times N$ matrix, $q$ is a constant unit quaternion
and the decomposition into blocks
is as in equation (\ref{adhmop}). The transformation rotates the
components of the vector $\Psi$, as can be seen from its definition
(\ref{adhmlin}), but this does
not change the gauge potential derived from the formula
(\ref{adhmpot}). 

Symmetric instantons within the ADHM formulation are described in detail in
ref.\cite{SiSu} and we only recall the main aspects here.
We are interested
in instantons which are symmetric under the action of a finite rotation group
$\G\subset SO(3)$ 
acting on the coordinates $(x_1,x_2,x_3)$ of $\R^3\subset\R^4$
and leaving $x_4$ alone. 
The quaternionic representation of a point $x\in\R^4$ in the ADHM
construction means that 
it is convenient to work with the binary group $\tilde\G$,
which is the double cover of $\G$ obtained from the double
cover of $SO(3)$ by $SU(2)$.
 Now we can exploit the equivalence of
$SU(2)$ and the group of unit quaternions to represent an element of 
$\tilde\G$ by a
unit quaternion $g$, with spatial rotation acting by the conjugation
\be
x\mapsto gxg^{-1}\,,
\label{quatrot}
\ee
which fixes the $x_4$ component and transforms the pure part by
the $SO(3)$ rotation corresponding to the $SU(2)$ element represented by $g$.
The ADHM data of an $N$-instanton is $\G$-symmetric if for every $g\in
\tilde\G$ the spatial rotation (\ref{quatrot}) leads to gauge equivalent ADHM
data. Recalling the redundancy (\ref{adhmred}), the requirement is
that for every $g$
\be
\pmatrix{L \cr M-gxg^{-1}1_N}=
\pmatrix{q &  0 \cr 0 & {\cal O} g}
\pmatrix{L \cr M-x1_N}g^{-1}{\cal O}^{-1}\,,
\label{adhmsym}
\ee
where, as earlier, ${\cal O}\in O(N)$ and $q$ is a unit quaternion,
both being $g$-dependent. The set of matrices ${\cal O}(g)$, as $g$ runs
over all the elements of $\tilde\G$, forms a real $N$-dimensional
representation of $\tilde\G$, and similarly the set of quaternions $q(g)$
forms a quaternionic one-dimensional representation or 
equivalently a complex two-dimensional representation. 
The procedure to calculate
$\G$-symmetric ADHM data is therefore first to choose a real
$N$-dimensional representation of $\tilde\G$, which we shall denote by $W,$ 
and a complex two-dimensional
representation of $\tilde\G,$ which we shall denote by $Q,$
 and then to find the most general
 matrices $L$ and $M$
compatible with equation (\ref{adhmsym}). Hopefully, these matrices
then contain few enough parameters to make the ADHM constraint
(\ref{adhmcon}) tractable, yet non-trivial.

\newpage
\section{Representations of the Binary Icosahedral Group}
\news\label{sec-reps}\hskip 0.5cm
In this paper we are concerned with icosahedrally symmetric instantons,
so we shall require some details of the representation theory of
the binary icosahedral group $\tilde Y.$
There are nine irreducible representations of $\tilde Y,$ and these
are listed in Table \ref{tab-repY} together with their dimensions.
A prime on a representation denotes that it is not a representation
of $Y,$ but only of the binary group $\tilde Y.$
\begin{table}
\centering\begin{tabular}{|c||c|c|c|c|c|c|c|c|c|}\hline
irreps of $\tilde Y$ &$A$ &$E_1'$ &$E_2'$ &$F_1$ &$F_2$ &$G$ &$G'$ & $H$ &
$I'$\\ \hline
dimension &1 &2 &2 &3 &3 &4 &4 &5 &6  \\
\hline\end{tabular}
\caption{Irreducible representations of $\tilde Y$.}
\label{tab-repY}
\end{table}

The representations $A,E_1',F_1,G',H,I',$ of dimension $d=1,2,3,4,5,6$ 
are obtained as the restriction $\underline{d}|_{\tilde Y}$ of the
corresponding $d$-dimensional irreducible representation of $SU(2).$
As for the remaining representations, $E_2'$ and $F_2$ are 
obtained from the representations $E_1'$ and $F_1$ by making the 
replacement $\sqrt{5}\mapsto -\sqrt{5}$ in the character table,
and $G=E_1'\otimes E_2'.$ 

\newcommand{\iv}{^{-1}}

The binary icosahedral group is generated by the three 
unit quaternions \cite{Cox}
\be
g_1=i,\ \
g_2=j,\ \
g_3=-\frac{1}{2}(i+\tau j-\tau^{-1}k)
\label{gen3}\ee
where $\tau=\half(\sqrt{5}+1)$ is the golden mean.
This quaternionic one-dimensional representation corresponds
to the complex two-dimensional representation $E_1'.$

In the following section we shall require expressions for 
these three generators in the representations $E_2',F_2,G$ and $H,$
so we present them here.

Regarding $E_2'$ as a one-dimensional quaternionic representation 
the three generators are obtained by 
making the replacement $\tau\mapsto -\tau^{-1},$ 
in the expressions (\ref{gen3})
\be
q(g_1)=i,\ \
q(g_2)=j,\ \
q(g_3)=-\frac{1}{2}(i-\tau^{-1} j+\tau k)
\label{repE2'}
\ee
this corresponds to the replacement $\sqrt{5}\mapsto -\sqrt{5}$ mentioned
above.

In $F_2$ they are represented
by
$$
{\cal O}_{F_2}(g_1)=\pmatrix{1 & 0 & 0\cr 0 & -1 & 0 \cr 0 & 0 & -1\cr}\,,\ \
{\cal O}_{F_2}(g_2)=\pmatrix{-1 & 0 & 0\cr 0 & 1 & 0 \cr 0 & 0 & -1\cr},$$
\be {\cal O}_{F_2}(g_3)=-\frac{1}{2}
\pmatrix{ 1 & \tau\iv & -\tau\cr \tau\iv & \tau & 1\cr
-\tau & 1 & -\tau\iv\cr},
\label{repF2}\ee
and in $G$ they are
$$
{\cal O}_{G}(g_1)=\pmatrix{1 & 0 & 0 & 0\cr
                       0 & 1 & 0 & 0\cr
                       0 & 0 & -1 & 0\cr
                       0 & 0 & 0 & -1\cr}\,, \ \
{\cal O}_G(g_2)=\pmatrix{1 & 0 & 0 & 0\cr
                       0 & -1 & 0 & 0\cr
                       0 & 0 & 1 & 0\cr
                       0 & 0 & 0 & -1\cr},\ $$
\be 
{\cal O}_G(g_3)=\frac{1}{4}\pmatrix{-1 &  \sqrt{5} &- \sqrt{5} & -\sqrt{5}\cr
\sqrt{5} & 3 & 1 & 1\cr
-\sqrt{5} & 1 & -1 & 3\cr -\sqrt{5} & 1 & 3 & -1\cr}.
\label{repG}\ee
Finally, in $H$ they are given by
$$
{\cal O}_H(g_1)=\pmatrix{1 & 0 & 0 & 0 & 0\cr
0 &1 & 0 & 0&0\cr
0&0&1&0&0\cr
0&0&0&-1&0\cr
0&0&0&0&-1\cr}\,, \\
{\cal O}_H(g_2)=\pmatrix{1 & 0 & 0 & 0 & 0\cr
0 &-1 & 0 & 0&0\cr
0&0&1&0&0\cr
0&0&0&-1&0\cr
0&0&0&0&1\cr}\,,$$
\be
{\cal O}_H(g_3)=\frac{1}{4}\pmatrix{
-1 & \sqrt{2} & -\sqrt{3} & \sqrt{2} & -\sqrt{8}\cr
\sqrt{2} & 0 & -\sqrt{6} & 2 &2\cr
-\sqrt{3} & -\sqrt{6} & 1 &\sqrt{6} & 0\cr
\sqrt{2} & 2 & \sqrt{6} & 2 & 0\cr
-\sqrt{8} & 2 & 0 & 0 &2\cr}.
\label{repH} \ee

\section{ADHM Data for the Buckyball}\news\label{sec-bucky}\hskip 0.5cm
The first step in attempting to construct an icosahedrally
symmetric charge seventeen instanton is to choose $W,$ the real 17-dimensional
representation of $\tilde Y.$ Studies of symmetric monopoles 
\cite{HMM,HS1,HS2} suggests that 
when searching for $\tilde\G$-symmetric instantons
a fruitful choice for the $N$-dimensional
space $W$ is the restriction of the $N$-dimensional irreducible
representation of $SU(2)$ ie.
\be
W=\underline{N}|_{\tilde\G}.
\ee 
Making this choice with $N=17$ and $\tilde\G=\tilde Y$ gives
\be
W=\underline{17}|_{\tilde Y}=F_2\oplus G\oplus 2H,
\ee
which explains why we presented the details for these real representations
in the previous section.

From equation (\ref{adhmsym}) we see that
\be
q(g)L{\cal O}(g)^{-1}g^{-1}=L
\ee 
for all $g\in\tilde Y.$ This equation means that $L$ is a 
$\tilde Y$-invariant map from $W\otimes E_1'$ to $Q.$ Now since 
\be
W\otimes E_1'=(F_2\oplus G\oplus 2H)\otimes E_1'=
I'\oplus (E_2'+I') \oplus 2(G'+I')
\ee
then we must have that $Q=E_2',$ since this is the only two-dimensional
representation that occurs in the final expression above.
To find a basis, say $L_1,$ for the invariant map $G\otimes E_1'\mapsto E_2'$
the quaternionic linear equations
\be
q(g_s)L_1{\cal O}_G(g_s)^{-1}g_s^{-1}=L_1
\ee
must be solved for $L_1$ with $s=1,2,3,$ 
where the generators $g_s$ are given by
(\ref{gen3}), and the representation of the generators $q(g_s)$ in
$E_2'$ and ${\cal O}_G(g_s)$ in $G$ are given in (\ref{repE2'})
and (\ref{repG}) respectively. These quaternionic linear equations, and
all similar equations later in the paper, were solved using {\rm{MAPLE}}
with the quaternions dealt with using the Clifford algebra package
{\rm{CLIFFORD}} \cite{clifford}. The result is that
\be
L_1=(1,i,j,k)
\ee
with any real multiple of $L_1$ being the general invariant map.

Equation (\ref{adhmsym}) reveals that for all $g\in\tilde Y$ 
\be
{\cal O}(g)gM{\cal O}(g)^{-1}g^{-1}=M
\ee
which implies that we may view $M$ as a $\tilde Y$-invariant map from $W$ to
$W\otimes E_1'\otimes E_1'.$ Now $E_1'\otimes E_1'=A\oplus F_1$ and this
corresponds to the decomposition of $M$ into a real and pure quaternion
part. The real part gives a multiple of the identity matrix for each
irreducible component of $W$ and to compute the pure part
we must construct the general invariant map $W\mapsto W\otimes F_1.$ 

From the following products of representations
\be
F_2\otimes F_1=G\oplus H\,,\ \
G\otimes F_1 = F_2\oplus G\oplus H\,,\ \
H\otimes F_1=F_1\oplus F_2\oplus G\oplus H
\ee
we see that the pure part of $M$ must be constructed from the invariant maps
\bea
B_1: & & F_2 \mapsto G\otimes F_1\label{B1}\\
B_2: & & F_2 \mapsto H\otimes F_1\\
B_3: & & G \mapsto H\otimes F_1\\
B_4: & & G \mapsto G\otimes F_1\\
B_5: & & H \mapsto H\otimes F_1\\
B_1^\dagger: & & G \mapsto F_2\otimes F_1\\
B_2^\dagger: & & H \mapsto F_2\otimes F_1\\
B_3^\dagger: & & H \mapsto G\otimes F_1\label{B3d}.
\eea
To obtain a basis for each of these maps, let $B$ denote one of the above
maps such that $B: R_2 \mapsto R_1\otimes F_1,$ 
where $R_1$ and $R_2$ each denote 
one of the representations $F_2, G$ or $H$. Then $B$ is the pure quaternion
matrix of dimension ${\rm dim} R_1 \times {\rm dim} R_2$ that solves the
quaternionic linear equations
\be
{\cal O}_{R_1}(g_s)g_sB{\cal O}_{R_2}(g_s)^{-1}g_s^{-1}=B
\ee
with $s=1,2,3.$ Using the explicit matrices given in section \ref{sec-reps}
these equations can be solved using {\rm {MAPLE}} to yield 
\be
B_1=\pmatrix{
i&j&k\cr
0&\tau k& \tau^{-1} j\cr
\tau^{-1}k & 0 & \tau i\cr
\tau j & \tau^{-1}i & 0},\
B_2=\pmatrix{
i&j&-2k\cr
0 & -\sqrt{2}\ti k & \sqrt{2}\tau j\cr
-\sqrt{3} i & \sqrt{3} j & 0\cr
\sqrt{2}\ti j & -\sqrt{2}\tau i & 0\cr
-\sqrt{2}\tau k & 0 & \sqrt{2}\ti i\cr}  
\label{defB1B2}\ee
\be
B_3=\pmatrix{
0 & (1-3\sqrt{5})i & (1+3\sqrt{5})j & -2k\cr
-2\sqrt{10}i & 0 & \sqrt{2}(3+\sqrt{5})k & \sqrt{2}(3-\sqrt{5})j\cr
0 & -\sqrt{3}(1+\sqrt{5})i& \sqrt{3}(1-\sqrt{5})j& 2\sqrt{15}k\cr
 2\sqrt{10}k & -\sqrt{2}(3+\sqrt{5})j & \sqrt{2}(3-\sqrt{5})i& 0\cr
 2\sqrt{10}j & -\sqrt{2}(3-\sqrt{5})k & 0 & -\sqrt{2}(3+\sqrt{5})i\cr}.
\label{defB3}\ee
Note that $B_1B_1^\dagger$ is an invariant map 
$B_1B_1^\dagger: G\mapsto G\otimes E_1'\otimes E_1',$ so its
pure quaternion part is a basis for the map $B_4$ 
\be
B_4={\rm{Im}}(B_1B_1^\dagger)=\pmatrix{
0 & -i & -j & -k\cr
i & 0 & k & -j\cr
j&-k&0&i\cr
k&j&-i&0\cr}
\label{defB4}\ee
where ${\rm{Im}}$ denotes the pure quaternion part.

\noindent Similarly, $B_2B_2^\dagger: H\mapsto H\otimes E_1'\otimes E_1',$
 so its pure quaternion part is a basis for the map $B_5$ 
\bea
B_5&=&{\rm{Im}}(B_2B_2^\dagger)\label{defB5}\\
&=&\frac{1}{\sqrt{2}}\pmatrix{
0 & -(\sqrt{5}+3)i & 0 & -2\sqrt{5}k & (\sqrt{5}-3)j\cr
(\sqrt{5}+3)i & 0& -\sqrt{3}(\sqrt{5}-1)i & -2\sqrt{2}j & 2\sqrt{2}k\cr
0 & \sqrt{3}(\sqrt{5}-1)i & 0 & -2\sqrt{3}k & \sqrt{3}(\sqrt{5}+1)j\cr
2\sqrt{5}k & 2\sqrt{2}j & 2\sqrt{3}k & 0 & 2\sqrt{2}i\cr
-(\sqrt{5}-3)j & -2\sqrt{2}k & -\sqrt{3}(\sqrt{5}+1)j & -2\sqrt{2}i & 0\cr}.
\nonumber \eea

Note that the nature of the above construction for 
$B_4$ and $B_5$ means that $B_4=B_4^\dagger$ and  $B_5=B_5^\dagger.$

The matrices $B_1,...,B_5$, and their quaternionic conjugates, 
together with the identity matrices,
are a basis for all the invariant maps between the spaces we are
considering, so the (allowed) products of any two can be written as a
linear combination of this set. Using the explicit matrices listed above
we compute the following product formulae that are required later
\bea
& &B_1B_1^\dagger=3\, 1_4  +B_4,\
B_2B_1^\dagger=\half B_3,\
B_2B_2^\dagger=6\, 1_5 +B_5,\
B_3B_3^\dagger=96\, 1_5 +16B_5,\ \nonumber\\
& &B_3B_1=8B_2,\ 
B_5B_5=24\, 1_5-2B_5,\
B_5B_2=4B_2,\
B_5B_3=4B_3,\ \label{prods}\\
& &B_1^\dagger B_1=4\, 1_3,\
B_2^\dagger B_2=10\, 1_3,\ 
B_3^\dagger B_3=120\, 1_4 +40B_4,\
B_3^\dagger B_2=20B_1.\nonumber
\eea
As $L_1$ is also an invariant map, we find that $L_1^\dagger L_1=1_4-B_4.$

As none of the $B$ matrices are symmetric, they can only be assembled to
form the symmetric matrix $M$ if they are placed in off-diagonal blocks.
As all the $B$ matrices are pure quaternion then $B^\dagger=-B^t,$ and
this determines the block structure of $\widehat M$ to be
\be
\widehat M=\lambda \pmatrix{
0 & L_1 & 0 & 0\cr
\beta_11_3 & -\alpha_1 B_1^\dagger & 
-\alpha_2 B_2^\dagger & -\alpha_3 B_2^\dagger\cr
\alpha_1 B_1 & \beta_2 1_4 &
-\alpha_6 B_3^\dagger & -\alpha_4 B_3^\dagger\cr
\alpha_2 B_2 & \alpha_6 B_3 & \beta_3 1_5 & -\alpha_5 B_5\cr
\alpha_3 B_2 & \alpha_4 B_3 & \alpha_5 B_5 & \beta_4 1_5\cr}
\label{blocks}
\ee
where $\alpha_1,..,\alpha_6,\beta_1,...,\beta_4$ are real constants
and $\lambda$ is an arbitrary non-zero real constant which
 sets the overall scale of the instanton. 
We fix the instanton scale by choosing $\lambda=1$ from now on.

The invariant map (\ref{blocks}) must now be subjected to the ADHM constraint
(\ref{adhmcon}). Computing the product
$\widehat M^\dagger\widehat M$ produces a block form in which each 
block is proportional to one of the $B$ matrices plus a possible 
contribution proportional to an identity matrix. To satisfy the
ADHM constraint all the terms proportional to the $B$ matrices must
vanish. Applying the product formulae (\ref{prods}) yields the
equations
\bea
&&\alpha_1(\beta_2-\beta_1)+20\alpha_2\alpha_6+20\alpha_3\alpha_4=0\nonumber\\
&&\alpha_2(\beta_3-\beta_1)-8\alpha_1\alpha_6+4\alpha_3\alpha_5=0\nonumber\\
&&\alpha_3(\beta_4-\beta_1)-8\alpha_1\alpha_4-4\alpha_2\alpha_5=0\nonumber\\
&&\alpha_4(\beta_4-\beta_2)+\half\alpha_1\alpha_3-4\alpha_5\alpha_6=0\nonumber\\
&&\alpha_5(\beta_4-\beta_3)+\alpha_2\alpha_3+16\alpha_4\alpha_6=0\nonumber\\
&&\alpha_6(\beta_3-\beta_2)+\half\alpha_1\alpha_2+4\alpha_4\alpha_5=0\nonumber\\
&&\alpha_1^2+40\alpha_4^2+40\alpha_6^2=1\nonumber\\
&&\alpha_2^2+16\alpha_6^2=2\alpha_5^2\nonumber\\
&&\alpha_3^2+16\alpha_4^2=2\alpha_5^2.\label{cons}
\eea
These equations require that $\beta_1=\beta_2=\beta_3=\beta_4,$ and
hence the freedom in the arbitrary parameter $\beta_1$ simply corresponds
to a translation of the instanton in the $x_4$ direction. We fix this
freedom by setting $\beta_1=\beta_2=\beta_3=\beta_4=0.$

Note that there is a degenerate solution $\alpha_1=1,$ $\alpha_s=0$ for $s>1,$
for which only the first $8\times 7$ block of $\widehat M$ 
contains non-zero entries. This is the ADHM data of the icosahedrally
symmetric charge seven instanton found in \cite{SiSu}, for which the
topological charge density, at fixed Euclidean time, is localized on
the edges of an icosahedron. The similar solution with $\alpha_1=-1$
gives equivalent data.

The general solution (upto some sign changes which give equivalent data)
of the equations (\ref{cons}) is given by
\be
\alpha_1=\frac{2}{3},\
\alpha_2=\frac{\sqrt{2}}{3}\sin\theta,\
\alpha_3=\frac{\sqrt{2}}{3}\cos\theta,\
\alpha_4=-\frac{1}{6\sqrt{2}}\sin\theta,\
\alpha_5=\frac{1}{3},\
\alpha_6=\frac{1}{6\sqrt{2}}\cos\theta\
\ee
where $\theta$ is an arbitrary angle. In fact this whole one-parameter
family gives equivalent data, corresponding to a freedom to rotate
the $B_2$ and $B_3$ blocks inside $\widehat M.$ We can therefore choose
a convenient member of this family, $\theta=0,$ to give the solution
\be
\alpha_1=\frac{2}{3},\
\alpha_2=0,\
\alpha_3=\frac{\sqrt{2}}{3},\
\alpha_4=0,\
\alpha_5=\frac{1}{3},\
\alpha_6=\frac{1}{6\sqrt{2}}.
\ee
So finally, the ADHM data for the icosahedrally symmetric 17-instanton, 
which is unique upto the obvious freedom to scale, rotate and translate,
is given by
\be
\widehat M=\frac{1}{3}
\pmatrix{
0 & 3L_1 & 0 & 0\cr
0 & 2B_1^t & 0 & \sqrt{2}B_2^t\cr
2B_1 & 0 & \frac{1}{2\sqrt{2}}B_3^t & 0\cr
0 & \frac{1}{2\sqrt{2}}B_3 & 0 & B_5^t\cr
\sqrt{2}B_2 & 0 & B_5 & 0\cr
}.
\label{M17}
\ee

Given the explicit matrix (\ref{M17}) the real matrix $R(x)$,
defined by (\ref{rmatrix}), can be computed explicitly using
{\rm MAPLE}, and its determinant calculated to verify that
it is non-zero. Using the formula (\ref{neat}) a {\rm MAPLE}
computation can generate an explicit expression for the 
topological charge density, but this is such a horrendous expression
that it is not even efficient to use it to plot a topological
charge density isosurface. In fact a much more efficient numerical
scheme is to compute the determinant of the matrix $R(x)$ numerically
and use a finite difference approximation to the derivatives in
equation (\ref{neat}) to produce data for a plot. The results of
this scheme are displayed in Fig.~1, where we present a topological
charge density isosurface in $\R^3\subset \R^4,$ obtained at
zero Euclidean time $x_4=0.$ It can be seen that
the topological charge density is localized around the ninety edges
(and particularly the sixty vertices)
of the truncated icosahedron of the buckyball, as predicted.
As the only $x_4$ dependence of the matrix $R(x)$ is in the combination
$|x|^2=x_1^2+x_2^2+x_3^2+x_4^2,$ then isosurfaces for different Euclidean
time slices are qualitatively similar, though the level set value needs
to be reduced to correspond to the fact that the topological charge density
decreases as $x_4^2$ increases.

\begin{figure}[ht]
\begin{center}
\leavevmode
\vskip -0cm
\epsfxsize=7cm\epsffile{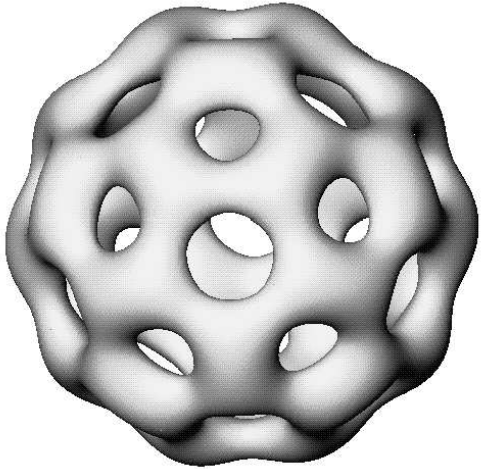}
\caption{A topological charge density isosurface, in the Euclidean
time slice $x_4=0,$ for the charge seventeen buckyball instanton.}
\label{fig-Y17}
\end{center}
\end{figure}

For a suitable choice of scale, the holonomy of this instanton will provide
 a good approximation to the minimal energy charge seventeen Skyrmion.
However, we have not investigated the energy of the resulting Skyrme
field to determine the required scale since it is computationally expensive 
and a good approximation to
this Skyrmion has already been obtained using a different approach \cite{HMS}.

\section{Conclusion}\hskip 0.5cm
The ADHM data has been obtained for an icosahedrally symmetric
charge seventeen instanton with a buckyball structure. 
The existence of this instanton was predicted by studying Skyrmions,
and this approach also predicts the existence of a whole range of
fullerene instantons. However, it is not clear which fullerenes
correspond to tractable ADHM data. There is evidence \cite{BHS} that
the minimal energy fullerene Skyrmion has icosahedral symmetry for charges in
the sequence which begins $7,17,37,67,97,...$ As we have seen, icosahedral
ADHM data is tractable for the first two charges in this sequence, so
it may be tractable for others too.

\section*{Acknowledgements}
I thank the EPSRC for an advanced fellowship.
\\


\begin{thebibliography}{99}

\bibitem{ADHM} M.F. Atiyah, V.G. Drinfeld, N.J. Hitchin and Yu.I. Manin,
Phys. Lett. A 65, 185 (1978).

\bibitem{AM} M.F. Atiyah and N.S. Manton, Phys. Lett. B 222, 438 (1989);
Commun. Math. Phys. 153, 391 (1993).

\bibitem{BHS} R.A. Battye, C.J. Houghton and P.M. Sutcliffe,
J. Math. Phys. 44, 3543 (2003).

\bibitem{BS2} R.A. Battye and P.M. Sutcliffe, Phys. Rev. Lett. 79, 363 (1997).

\bibitem{BS} R.A. Battye and P.M. Sutcliffe,
Phys. Rev. Lett. 86, 3989 (2001);
Rev. Math. Phys. 14, 29 (2002). 

\bibitem{BTC} E. Braaten, S. Townsend and L. Carson,
Phys. Lett. B 235, 147 (1990).

\bibitem{CWS} N.H. Christ, E.J. Weinberg and N.K. Stanton,
Phys. Rev. D 18, 2013 (1978).

\bibitem{CFGT} E. Corrigan, D.B. Fairlie, P. Goddard  and S. Templeton,
Nucl. Phys. B 140, 31 (1978).

\bibitem{Cox} H.S.M. Coxeter, {\sl Regular Complex Polytopes},
Cambridge University Press (1974).

\bibitem{HMM} N.J. Hitchin, N.S. Manton and M.K. Murray,
 Nonlinearity 8, 661 (1995).

\bibitem{HMS} C.J. Houghton, N.S. Manton and P.M. Sutcliffe,
Nucl. Phys. B {510}, 507 (1998).

\bibitem{HS1} C.J. Houghton and P.M. Sutcliffe, Commun. Math. Phys.
180, 343 (1996).
 
\bibitem{HS2} C.J. Houghton and P.M. Sutcliffe, Nonlinearity 9, 385 (1996).

\bibitem{JNR} R. Jackiw, C. Nohl and C. Rebbi, Phys. Rev. D 15, 1642 (1977).

\bibitem{LM} R.A. Leese and N.S. Manton, Nucl. Phys. A 572, 675 (1994).

\bibitem{Os} H. Osborn, Nucl. Phys. B 140, 45 (1978).

\bibitem{SiSu} M.A. Singer and P.M. Sutcliffe, Nonlinearity 12, 987 (1999).  

\bibitem{clifford} http://math.tntech.edu/rafal

\end{thebibliography}
\end{document}